\documentclass[sigconf,nonacm]{acmart}
\AtBeginDocument{%
  }

\setcopyright{acmlicensed}
\copyrightyear{2018}
\acmYear{2018}
\acmDOI{XXXXXXX.XXXXXXX}
\acmConference[Conference acronym 'XX]{Make sure to enter the correct
  conference title from your rights confirmation email}{June 03--05,
  2018}{Woodstock, NY}
\acmISBN{978-1-4503-XXXX-X/2018/06}

\newtoggle{comments}
\toggletrue{comments}

\iftoggle{comments} {
\newcommand{\lev}[1]{{\color{red}{#1}\normalfont}}
}{
  \newcommand {\lev}[1]{}
 }

\usepackage{xcolor}  %

\newcommand{\ws}[1]{\colorbox{yellow!30}{#1}}

\usepackage{xcolor}
\usepackage{tcolorbox}

\usepackage{multicol}
\usepackage{colortbl}
\usepackage{multirow}
\usepackage{hhline}
\usepackage{tabularx}
\usepackage{booktabs}
\newcolumntype{L}[1]{>{\raggedright\arraybackslash}p{#1}}

\begin{document}

\title{Understanding, Protecting, and Augmenting Human Cognition with Generative AI: \\A Synthesis of the CHI 2025 \textit{Tools for Thought Workshop}}

\author{Lev Tankelevitch}
\email{lev.tankelevitch@microsoft.com}
\affiliation{%
  \institution{Microsoft Research}
  \city{Cambridge}
  \country{United Kingdom}
}

\author{Elena L. Glassman}
\email{glassman@seas.harvard.edu}
\affiliation{%
  \institution{Harvard University}
  \country{USA}
}

\author{Jessica He}
\email{jessicahe@ibm.com}
\affiliation{%
  \institution{IBM Research}
  \country{USA}
}

\author{Aniket Kittur}
\email{nkittur@andrew.cmu.edu}
\affiliation{%
  \institution{Carnegie Mellon University}
  \country{USA}
}

\author{Mina Lee}
\email{mnlee@cs.uchicago.edu}
\affiliation{%
  \institution{University of Chicago}
  \country{USA}
}

\author{Srishti Palani}
\email{srishti.palani@salesforce.com}
\affiliation{%
  \institution{Tableau Research}
  \country{USA}
}

\author{Advait Sarkar}
\email{advait@microsoft.com}
\affiliation{%
  \institution{Microsoft Research}
  \city{Cambridge}
  \country{United Kingdom}
}

\author{Gonzalo Ramos}
\email{goramos@microsoft.com}
\affiliation{%
  \institution{Microsoft Research}
  \city{Redmond}
  \country{USA}
}

\author{Yvonne Rogers}
\email{y.rogers@ucl.ac.uk}
\affiliation{%
  \institution{University College London}
  \city{London}
  \country{United Kingdom}
}

\author{Hari Subramonyam}
\email{harihars@stanford.edu}
\affiliation{%
  \institution{Stanford University}
  \country{USA}
}

\renewcommand{\shortauthors}{Tankelevitch et al.}

\begin{abstract}
Generative AI (GenAI) radically expands the scope and capability of automation for work, education, and everyday tasks, a transformation posing both risks and opportunities for human cognition. How will human cognition change, and what opportunities are there for GenAI to augment it? Which theories, metrics, and other tools are needed to address these questions? The CHI 2025 workshop on Tools for Thought aimed to bridge an emerging science of how the use of GenAI affects human thought, from metacognition to critical thinking, memory, and creativity, with an emerging design practice for building GenAI tools that both protect and augment human thought. Fifty-six researchers, designers, and thinkers from across disciplines as well as industry and academia, along with 34 papers and portfolios, seeded a day of discussion, ideation, and community-building. We synthesize this material here to begin mapping the space of research and design opportunities and to catalyze a multidisciplinary community around this pressing area of research.
\end{abstract}


\begin{teaserfigure}
  \includegraphics[width=\textwidth]{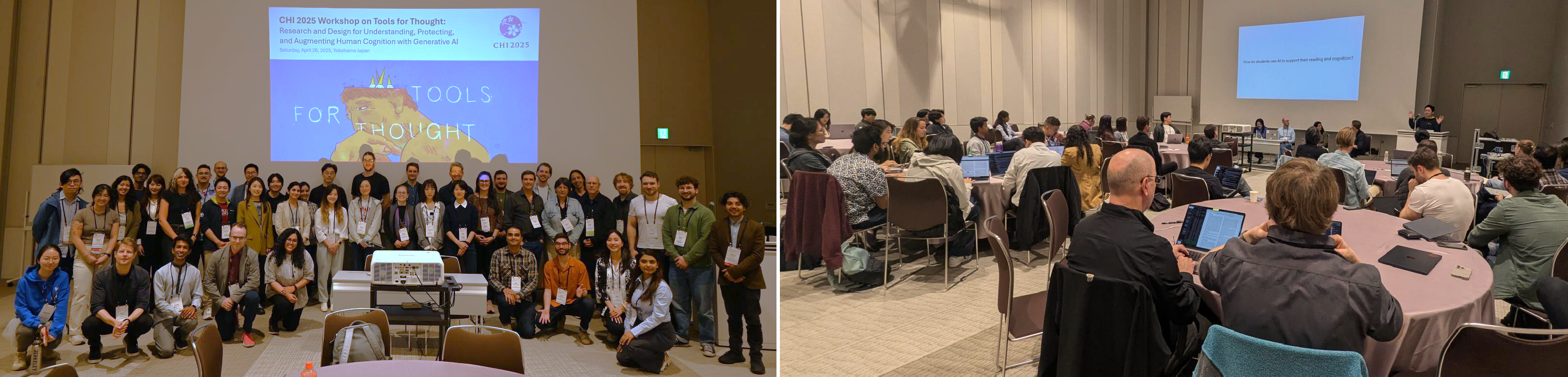}
  \caption{The CHI 2025 \textit{Tools for Thought Workshop} was held in Yokohama, Japan, on April 26th, 2025.}
  \Description{}
  \label{fig:teaser}
\end{teaserfigure}

\maketitle

\section{Introduction}
With the development of Generative AI (GenAI), society is undergoing a radical expansion in the scope and capability of automation and mechanization of cognitive work, a transformation which poses important risks for human cognition. Yet, the current moment also presents an expansion in the opportunities to \textit{positively} transform how we think. How will people learn and think in the future, as AI is embedded into our work and lives as a `co-worker', tool, or ubiquitous capability? How can we protect human cognition from the potential negative impacts of AI-driven automation? What opportunities are there for AI to augment thinking, much as older external tools such as writing have done? Which theories, metrics, and other tools can best equip researchers and designers to address these questions? The CHI 2025 workshop on \textit{Tools for Thought} \cite{tankelevitch_tools_2025}---systems or methods that extend, support, or augment cognition \cite{engelbart1962augmenting, victor2013media}---aimed to bridge an emerging science of how the use of AI affects human thought, from metacognition to critical thinking, memory, and creativity, with an emerging design practice for building AI tools that not only protect human thought but also augment it. Concretely, it aimed to map the space of research and design opportunities in this area, and catalyze a multidisciplinary community interested in pursuing them. 

Held in Yokohama in April 2025, the workshop convened 56 researchers, designers, and thinkers from industry and academia, covering fields such as human-computer interaction and other aspects of computer science, cognitive science, design, and education. Thirty-four accepted papers and portfolios (selected from over 70 submissions) set the stage for a day of discussion, ideation, and community-building.\footnote{Throughout the paper, each accepted submission is first introduced with a \ws{highlight like this.} Subsequent citations follow ordinary formatting conventions. PDFs of all accepted submissions can be found \href{https://ai-tools-for-thought.github.io/workshop/\#papers}{[here]}.} Here we synthesize the workshop outputs to advance the conversation on this pressing area of research. We structure this article around three areas: (i) \textit{understanding AI's impact on, and protecting, cognition}, (ii) \textit{augmenting human cognition with AI}, and (iii) \textit{formative research, theory, measurement, and evaluation}. %

\section{Understanding AI’s impact on, and protecting, human cognition}
AI systems are designed and commonly used to automate capabilities and processes such as writing and research, processes through which people think, and therefore learn, build skills, and grow expertise. The generative nature of these systems means that human workflows are shifting from production to ‘critical integration’ of material \cite{sarkar_exploring_2023}, i.e., decisions about when and how to use AI, how to frame tasks, and how to assess outputs. Thus, the quality and nature of thinking, such as where and what kind of thinking is applied, all change when knowledge workflows reorient themselves around AI \cite{tankelevitch_metacognitive_2024}. As the next sections illustrate, these changes may manifest in critical thinking, learning, and creativity, among other cognitive processes. They have implications for workflows and our notions of expertise, as well as motivation, well-being, and other human values. 

\subsection{Critical thinking}
As \ws{\citet{singh_protecting_2025}} observe in emerging research, AI may discourage people from engaging in critical thinking, as seen, for example, in the shift from active information seeking to more passive consumption of AI-generated information (e.g., see \cite{danry_deceptive_2025}). This is exacerbated by GenAI systems’ tendencies for agreeing with users' ideas and homogenizing outputs, features which may contribute to the known `echo chamber' effects in contemporary information consumption. Beyond mere consumption, people may also inappropriately rely on inaccurate or misleading AI-generated information in everyday decisions \cite{singh_protecting_2025}, particularly as they may disengage from critical thinking when their self-confidence in a domain is low, or when their confidence in AI is high \cite{lee_impact_2025}. For example, biased writing assistants can shift users’ views on key issues \cite{williams-ceci_biased_2025} and homogenize writing toward western styles \cite{agarwal_ai_2025}. More broadly, AI use has been shown to exert an aggregate influence on the intentions of knowledge workers \cite{sarkar2024intention}, nudging the outputs of groups of workers towards smaller, less diverse sets of ideas, in the process of ``mechanized convergence'' \cite{sarkar_exploring_2023}.

How can we understand these changes from the perspective of human cognition? \citet{singh_protecting_2025} point to Dewey’s theory of reflective thinking \cite{dewey_how_1910} as a powerful explanatory lens. Reflective thinking, as an intentional and conscious evaluation of evidence, requires enduring a `state of perplexity, confusion, or doubt prompting inquiry' \cite{singh_protecting_2025}, as well as a suspension of judgment during this period of inquiry. However, the polished, seemingly coherent, and sycophantic nature of AI-generated information may minimize this prerequisite state, discourage suspension of judgment, and create an `illusion of comprehensive understanding' \cite{singh_protecting_2025}. Empirically testing which aspects of AI output and interaction patterns causally impact critical thinking in different contexts is an important and ongoing research direction \cite{kimFosteringAppropriateReliance2025, kimImNotSure2024, lee_impact_2025,okoso_expressions_2025}. How these effects play out after prolonged use is another important question. Worryingly, the potential impacts on critical thinking may be particularly acute in a learning context.

\subsection{Learning} \label{subsec: learning}
Unlike experts, students or novices are still developing the self-regulation strategies and schemas (i.e., mental models or sense-making frames) necessary for information work \cite{siu_augmenting_2025, singh_protecting_2025, nishal2025designing}. Thus, in addition to AI-generated output that can discourage a reflective state, students can also lack a developed skill set for prompting and critically evaluating AI output. For example, in a study of AI-assisted ML code debugging, \ws{\citet{bo_whos_2025}} found that students who already had an understanding of the problem used AI intentionally by asking specific, planned questions, and ultimately performed better. In contrast, those who lacked such knowledge asked more open-ended questions and were led by the AI down unproductive paths, engaging in more over-reliance. As \ws{Prather and Reeves \cite{prather2025pedagogy}} conclude about AI assistance in CS education, ``underprepared, unconfident, and underperforming students seem to benefit the least from these [AI] tools, resulting in less critical thinking and lower grades''. Thus, those populations most likely to rely on AI due to existing performance disparities may benefit the least, or may even be harmed by using it. 

By inappropriately relying on (non-pedagogical) AI tools to help solve their learning tasks, students may also impede their development of necessary metacognitive strategies and schemas \cite{singh_protecting_2025}. Students access synthesized and tailor-made information without engaging in the cognitive processes typical of deliberate learning, e.g., remembering, understanding, and applying disparate sources of information, as outlined in Bloom’s revised taxonomy of educational goals \cite{krathwohl_revision_2002}. Indeed, in a study of AI-assisted reading, \ws{\citet{fu_supporting_2025}} find that students may increasingly offload lower-level understanding to AI, consistent with more passive engagement. These observations naturally raise questions such as: how do we help people use GenAI effectively for \textit{learning}, rather than, or alongside, productivity purposes? What can we gather from those already using GenAI successfully for this purpose? 

\subsection{Creativity}
In the creative domain, \ws{\citet{dalsgaard2025tools}} argues that GenAI's facility for producing higher-fidelity outputs risks shifting designers’ focus from understanding the core problem framing or design space to the mere tweaking of polished outputs, a form of design fixation \cite{lin_inkspire_2025}. There is also a risk of GenAI homogenizing creative outputs \cite{anderson_homogenization_2024,doshi_generative_2024}, an effect potentially amplified by the challenge of working effectively with GenAI systems and grasping their subtle limitations \cite{tankelevitch_metacognitive_2024} (see also \cite{bangerl_creaitive_2025,kumar_human_2025}). 

Yet, \citet{dalsgaard2025tools} positions GenAI not solely as a risk for creativity, but rather as providing `instruments of inquiry': a set of tools that shape and guide creative thinking in myriad ways. Grounded in pragmatist philosophy, %
this perspective suggests that GenAI has a role in shaping people's perception about a design problem, their conceptions, how they externalize and manipulate ideas, and the mediation of the components of a design task. The nature of these effects depends on users' approach and skill set for working with GenAI. For example, designers can leverage GenAI's ability for rapid, high-fidelity prototyping to better engage in Schön’s process of ‘reflection-in-action’ \cite{schon2017reflective}, where iterative interaction with a design material affords a rapid feedback loop. Likewise, designers aware of GenAI's limitations can consciously exploit these for more creative outcomes. Empirically disentangling helpful and harmful uses of GenAI in creative processes and workflows, and the roles of interaction patterns and system capabilities is an important research question (e.g., \cite{bangerl_creaitive_2025}).   

Aspects of GenAI systems, such as the dominance of text-based prompting or their limited explainability, also raise questions about how to preserve practitioners' flow states and integrate GenAI into existing practices \cite{palani_evolving_2024}. \ws{\citet{naqvi2025computational}} highlight Sawyer’s model of the creative process, which characterizes it according to eight features such as iteration, reflection, ambiguity, exploration, and emergence. GenAI may be shifting how these characteristics arise in creative workflows. For example, reflection may shift from a focus on one’s own creative process and output to prompting and evaluating AI-generated output, whereas ambiguity may now be driven more by models’ lack of explainability or other limitations, rather than the problem space or iterative workflow itself. \ws{\citet{quintana2025students}} find evidence of this in AI-assisted learning experience design, where attention may shift from a focus on the design workflow itself to prompt engineering AI to get the desired output. What do these workflow changes mean for students who are still developing the schemas necessary for design work? How can AI become a `ready-to-hand tool', available for almost unconscious use, rather than a `present-at-hand' tool that requires focused effort \cite{quintana2025students}?\footnote{A distinction proposed by Heidegger \cite{heidegger1927being}, noted by Winograd and Flores \cite{winograd1987understanding}, and echoed in the `Extended Mind' thesis \cite{clark_extended_1998}.}    

\subsection{Contextual and applied factors: Workflows, expertise, and human values}\label{subsec:contextualfactors}

Beyond the impact on any one cognitive domain, several contextual and applied factors are central to understanding how GenAI may impact cognition in realistic settings. These include considerations of workflows, expertise, and human values around intrinsic motivation, well-being, and other aspects.    

\subsubsection{Workflows}\label{subsec:workflows}

Many of the changes driven by GenAI have been described as a shift to cognitive or metacognitive ‘laziness’ \cite{fan_beware_2025}, yet such characterizations themselves require reflection. When does ‘lazy’ simply become ‘resource reallocation’? Is cognitive offloading to GenAI instead better viewed as a form of `cognitive miserliness' \cite{stanovich2009cognitive} or `satisficing' \cite{simon1956rational}? Are there points in a workflow that benefit from cognitive offloading such that conserved cognitive energy can be deployed to other tasks? \footnote{Conserved energy may equally be deployed across workflows, or to non-work tasks entirely, such as leisure activities, a topic broaching the role of human values, as per §\ref{subsubsec:values}.} Understanding \textit{workflows}---complex, non-linear combinations of discrete tasks---may be one key to answering these questions. In programming, for example, whereas code generation and debugging may be increasingly solved by GenAI assistance, might design specification, refactoring, and code management need to remain essential human skills \cite{prather2025pedagogy}? In design, using GenAI to rapidly generate many alternatives may facilitate ideation, while the task of sifting through, combining, and curating alternatives may remain a human endeavor \cite{dalsgaard2025tools}. Focusing on tasks at the expense of workflows risks missing interactions between tasks. For example, if GenAI is introduced too early in the ideation process, rather than augmenting ideation, it can stifle creativity by fixating users’ attention onto specific ideas \cite{qin_timing_2025,lin_inkspire_2025}. Moving beyond discrete tasks, how can we better understand workflows, and GenAI's ability to reshape them?

\subsubsection{The role of expertise}\label{subsubsec:expertise}

Interlinked with workflows is the role of people's \textit{expertise}, another key factor in understanding the impact of AI on cognition. As \ws{\citet{siu_augmenting_2025}} summarize: 

\begin{quote}
    \textit{``experts develop sophisticated mental models and perceptual skills that enable them to maintain flexible cognitive frames throughout their sense-making process, allowing them to recognize meaningful patterns and adapt their understanding as new information emerges''}
\end{quote}

Experts in a particular domain may therefore be better equipped to specify goals, decompose tasks, and ultimately prompt GenAI systems in that context. They are also better able to assess, interpret, and validate the varied outputs produced by GenAI \cite{siu_augmenting_2025}. As per §\ref{subsec: learning}, these differences are even observable among students, with more advanced students using GenAI in more effective ways \cite{bo_whos_2025,margulieux_self-regulation_2024,prather_widening_2024}. 

Yet, although experts in domains such as research \cite{siu_augmenting_2025} and journalism \cite{nishal2025designing} may be keen to offload routine, low-level tasks to AI, they hesitate to offload higher-level analysis and decision-making to systems that they suspect lack the necessary nuance and context. %
Partly, this hesitation may be explained by the limitations in the accuracy of current GenAI systems, particularly for the kind of complex work in which experts engage \cite{jaffe_generative_2024, simkute_ironies_2024,becker_measuring_2025}. Part of it may also lie in workflow compatibility. Experts often have existing workflows and styles that they want AI to augment, rather than override \cite{baxter_ironies_2012}. This raises questions such as: how can we design AI systems that enable selective delegation, preserve agency, and support verification in a manner that respects varying levels and types of expertise \cite{siu_augmenting_2025}?  

A provocation here is whether experts are always better suited to judging when and what to offload to GenAI. One critical stance suggests that experts’ extensive accumulation of experience can lead to cognitive entrenchment that misses alternative or innovative approaches \cite{siu_augmenting_2025}. Is optimal offloading to GenAI therefore an emerging kind of expertise, or meta-expertise, one that involves delegation, navigating multiple domains, and orchestrating tools across workflows \cite{tankelevitch_metacognitive_2024}? AI-assisted coding offers a model: success here depends on task decomposition, evaluation, and preference, not just rote instruction. More broadly, this hints at the fact that expertise comes in different flavours: domain expertise differs from expertise with AI, which also differs from managerial expertise (i.e., the ability to define and delegate tasks to others). How do we better understand the types of expertise involved in working optimally with GenAI, including delegation, orchestration, and navigation across multiple domains and workflows? Acknowledging and supporting different types of expertise will be crucial for effective use of AI.      

An equally important question is how the use of GenAI and its focus on efficiency may change experts’ experience of their work, particularly reducing the space for `deep work'---i.e., a distraction-free focus on cognitively demanding and meaningful tasks, aimed to achieve meaningful progress \cite{newport2016deep}. For example, as \ws{\citet{nishal2025designing}} have found, journalists emphasize the importance of developing one’s writerly voice and the freedom to explore ideas, both of which may be curtailed by an AI-driven focus on productivity. How do we protect and incentivize time for `deep work' \cite{newport2016deep}? This raises broader considerations of motivation, well-being and other human values that sit alongside the impact on thinking and learning per se.

\subsubsection{Motivation, well-being, and other human values}\label{subsubsec:values}

How will GenAI, including its capability to automate cognitive tasks, affect our well-being? %
Citing Self-Determination Theory, \ws{\citet{langer2025why}} argues that ongoing human engagement in epistemic labor, i.e., actively acquiring, processing, and evaluating knowledge, is vital for sustaining intrinsic motivation and psychological well-being. Such epistemic labor fulfills the basic human needs for autonomy, competence, and relatedness, all of which may be threatened by excessive dependence on GenAI tools. Hence, the perceived sacredness of a writer’s voice \cite{nishal2025designing} or an artist’s style and its centrality to professional and personal identity, and attendant societal phenomena that discourage AI use on these grounds \cite{sarkar2025aishaming}. How does using AI affect people’s intrinsic motivation for tasks and other aspects of well-being \cite{scheiber_at_2025, schecter_how_2025, he_which_2025}?

This ties into the notion of `dialectical activities', pursuits such as parenting and art, that are valued for their intrinsic nature and sustained engagement \cite{zhang_searching_2024}. In creative domains, where taste and intentionality matter, the perceived interchangeability of GenAI outputs can feel particularly demotivating. For instance, journalists may appreciate the assistive value of GenAI, yet strongly resist AI generating the core ideas of their work, a perceived degradation of journalism’s artistic value \cite{nishal2025designing}. Yet, individuals differ in which tasks they find intrinsically motivating. Some may approach their work like sculptors, directly shaping material, while others may see themselves as miners, sifting through material in search of valuable insights. GenAI’s ability to produce material and reshape workflows carries distinct implications for each of these working styles which we need to understand \cite{sarkar_exploring_2023}. 

Lastly, considerations pertinent to well-being such as emotional vulnerability, dependency, and mental health in the context of GenAI interactions were only briefly touched upon at the workshop, yet are emerging as important issues \cite{fang_how_2025, devrio_taxonomy_2025, adam_supportive_2025}.

\section{Augmenting human cognition with AI}

By understanding the interplay between GenAI and human cognition, we can not only mitigate any potential negative impacts of GenAI, but also design systems that ultimately augment our thinking, leading to better cognitive and workflow outcomes in the short- and long- term. 

As evidenced by the range of our workshop submissions, approaches to GenAI-driven cognitive augmentation have taken, and should take, many forms. One dimension along which to organize current approaches is the \textit{directness} by which AI systems are designed to augment our cognition. Sitting at one extreme are designs intended to directly challenge our thinking through provocation \cite{drosos_it_2025,castaneda_supporting_2025} or questioning \cite{cheung2025from,danry_dont_2023}, discussed below in §\ref{subsec:provoke}. Less directly, some systems may not explicitly challenge us, but rather provide formal structure or scaffolding for processes such as sense making \cite{wang2025schemex} and ideation \cite{yang_overload_2025} (§\ref{subsec:scaffold}). Other approaches include using AI to transform the representation of information across modalities, fidelities, and other dimensions that may alter our thinking in subtle but potentially powerful ways (§\ref{subsec:transform}). Finally, some approaches harness emotional and motivational pathways, or other `System 1' processes \cite{evans_dual_2017}, to indirectly augment aspects such as learning, decision making, and creativity (§\ref{subsec:system1}).   

A common thread across these approaches is their focus on what \ws{\citet{zhang_augmenting_2025}} refer to as \textit{process-oriented} support, where AI assists users in identifying and addressing challenges throughout a task, ultimately enabling them to solve it themselves, rather than attempting to solve it for them (i.e., task automation, or `end-to-end' support; see also \cite{xu_productive_2025}). A key distinction of process-oriented support is its focus on helping users reason \textit{forward} towards their task solution---thereby supporting human understanding, verification, and agency---rather than reason backward from an AI-generated solution, which removes the user from the process (i.e., reducing situational awareness \cite{simkute_ironies_2024}) and increases the risk of over-reliance \cite{reicherts_ai_2025}. \ws{\citet{khurana2025designing}} identify that a semi-automated (or `process-oriented') approach---automating routine steps, while also guiding users through more complex decision-making---is more appropriate when users are less familiar with an application or task, when a task involves complex decision making or creativity (rather than solely repetitive or straightforward steps), or when users have learning goals (rather than solely performance goals; see also \cite{khurana_it_2025}).  

The next sections explore the different approaches to augmenting cognition. Cutting across these approaches are considerations about interacting and interfacing with GenAI (§\ref{subsec:interacting}), as well as the overarching workflows that encompass discrete tasks (§\ref{subsec:augworkflows}).  

\subsection{Provoking thinking through challenge} \label{subsec:provoke}%
One form of cognitive augmentation applies metaphors for AI as an active entity that challenges us during tasks: provocateur \cite{sarkar_ai_2024, sarkar2024genAIcritical}, antagonist \cite{cai_antagonistic_2024}, coach \cite{hofman_sports_2023}; disrupting our default modes of thinking, and prompting reflection, exploration, and learning. In the design domain, for example, \ws{\citet{gmeiner2025designing}} explore the use of metacognitive support agents that ask reflective questions and proactively support task planning and decomposition during the AI-assisted design process. They find that such interventions have the potential to support intent formulation, problem exploration, and outcome evaluation (see also \cite{gmeiner_exploring_2025,gmeiner_exploring_2023}). \ws{\citet{castaneda_supporting_2025}} take this approach one level higher: using AI to support meta-decision making (i.e., deciding how to decide), specifically helping users to iteratively define, refine, and prioritize their decision criteria. They propose a set of design goals for such assistance, instantiated in their prototype, including using concrete decision options or examples as reference points for users, systematically varying these options to help users discover decision dimensions of interest, and enabling iteration on all aspects of a decision. 

\ws{\citet{cheung2025from}} proposes another metaphor for AI-driven cognitive augmentation: \textit{the ignorant co-learner}, a deliberately ‘artificially ignorant’ system that fosters ‘moments of uncertainty, dissonance, or pause that compel users to think critically and reflexively’. For example, systems may offer ‘multiple, conflicting perspectives or highlight areas of uncertainty’, or prompts that call attention to the interpretive process in information consumption. While appealing in theory, all such approaches raise the question of how we can \textit{motivate} people to engage with systems that deliberately question or otherwise slow them down. Indeed, Cheung argues that AI provocations `must be clearly linked to pedagogical goals such as critical thinking, metacognition, and epistemic agency'. To this end, \ws{\citet{ayyappan2025building}} explore the value of explicitly clarifying the role of AI assistance during learning, allowing students to toggle between \textit{`Tell me the answer'} and \textit{`Ask me a question'} modes of AI support.

Yet GenAI is also being rapidly deployed across workplaces, where the primary aim is \textit{productivity} rather than learning. In this context, how can systems demonstrate the value of `extra' cognitive work to justify the time and effort taken, and thereby build intrinsic motivation to engage with them? Or, to question the doctrine of `desirable difficulty' \cite{bjork1994institutional}, must reflection or critical thinking always be \textit{difficult} or otherwise high-friction to be meaningful? What is the appropriate level and timing of friction for a given task? 

\subsection{Structuring and supporting tasks} \label{subsec:scaffold}%
Not all cognitive augmentation is designed to challenge thinking or provide reflective friction. Indeed, several of the above examples combine reflective prompts or other provocations with scaffolding intended to guide users through task completion. Prime examples of this latter approach often harness the flexibility and pattern recognition abilities of large language models (LLMs) to identify and surface structure within data to support sense making (e.g., \cite{dang_corpusstudio_2025}). \ws{\citet{wang2025schemex}} develop a human-AI complementary workflow that augments human schema induction---the process of identifying patterns from examples, such as articles or videos, to support sense making and learning. AI clusters examples by similarities, extracts structural patterns within each cluster, and refines clusters by generating contrastive examples. Humans evaluate and refine the generated schemas in an iterative loop. \ws{\citet{yang_overload_2025}} target human design ideation. Their prototype uses AI to first organize ideation stimuli into a functional hierarchy (i.e., a mechanism tree), and then generate analogical cues for these stimuli---e.g., how might a mechanism in context A be used in context B---to support users in identifying novel applications across domains (see also \cite{kang_biospark_2025}). Finally, \ws{\citet{schnizer2025user}} offer a vision for a GenAI-driven system for data visualization workflows that assesses and adapts to users' expertise, and provides dynamic suggestions, guidance, and reflective prompts.       

These explorations raise pertinent questions for understanding the impact on cognition. In contrast to the risk of AI systems deterring users via friction (§\ref{subsec:provoke}), one key concern here is the risk of \textit{over}-scaffolding tasks, with distinct implications for different cognitive processes. During sense-making, for example, which degree or type of cognitive effort is necessary for users to be helpfully close to the data and which can be bypassed? During the creative process, how do we protect the delicate, initial period of human idea incubation from the undue influence of AI? How does AI support during the divergence phase of ideation impact convergence? What is the appropriate balance of AI support between these two phases of ideation? Here, we can turn to the learning sciences, which define `learning goals' as a principled way to designing appropriate scaffolding \cite{puntambekar_tools_2005}. Alongside deliberate learning, how might such goals be defined in a productivity context? 

Overall, there is an interesting design tension between approaches that induce cognitive friction and those that scaffold tasks---finding leverage points and `sweet spots' in AI systems that get the best of both worlds is likely a fruitful area of future study.

\subsection{Transforming information representations}  \label{subsec:transform}%

Stepping outside of specific tasks, one thread of work explores how AI can be used to dynamically and interactively transform information between different representations, thus contributing to a known form of cognitive enhancement \cite{suh_dynamic_2024}. \ws{\citet{liu2025bridging}} explore how transforming content across modalities---e.g., from speech to text, or from text to visuals---can augment cognitive workflows. For example, how might flexibly translating between textual and graphical modalities enable the manipulation of concepts, rather than words or pixels? Their work raises key research questions for this space, such as how to best visualize the evolving semantic space of spoken content, or how to balance semantic abstraction with familiarity to support cognition.

Rather than modalities, \ws{\citet{yen2025something}} explore using AI to transform information between levels of \textit{formality}. In particular, they explore an approach to semi-formal programming---the integration of formal representations (e.g., defined variables in a piece of code) with informal ones (e.g., sketches). They propose three principles that work together: loosening strictly typed programs (e.g., dynamically converting unknown variables into placeholders), gradually enriching informal representations as meaning emerges through user interaction and context, and, rather than pre-specifying everything in advance, using AI to dynamically interpret or query fuzzy attributes as needed. 

Echoing across these projects is a model of human-AI partnership proposed by \ws{\citet{suh2025next}} in which AI expands humans' multidimensional conceptual environments (spanning knowledge, representations, design possibilities, among other aspects); humans further expand these environments via creativity, critical thinking, and contextual understanding; and together, they explore these environments via interfaces and interactions that thoughtfully support human-AI collaboration.

GenAI has surfaced many questions about the malleability of text---its ability to be flexibly summarized, expanded, and re-written on demand---and the implications for information processing, knowledge, and understanding. What are the affordances of text and other media as malleable projections of knowledge? How can systems that consume and render text support users in manipulating it to suit their cognitive goals and preferences? Beyond text, how do we effectively represent information in different output modalities to augment cognition? 

\subsection{Augmenting emotions and other `System 1' processes} \label{subsec:system1}
An implicit assumption underlying a lot of work on GenAI-driven cognitive augmentation is the prioritization of conscious reflection, deliberate reasoning, and other processes commonly associated with `System 2' thinking \cite{evans_dual-process_2013}.  Can (or should) we augment people's `System 1' processes---e.g., emotions, heuristics---to progress users towards desired cognitive outcomes \cite{matuschak_how_2019}? This may be welcomed, and indeed feel natural, in the creative domain. For example, \ws{\citet{pilcher_purposefully_2025}} propose an approach to fostering creativity by welcoming and amplifying GenAI's inaccuracies or fabrications (`hallucinations'). Their mixed-reality prototype envelopes users in surreal and occasionally unsettling multi-modal outputs to provoke `surprise, reflection, and creative risk-taking'. 

What would augmenting `System 1' processes look like in other tasks? In learning, for example, \ws{\citet{leong2025designing}'s} work demonstrates how AI can help change users' perspectives through emotional and motivational pathways. For instance, AI can personalize experiences to users' interests, such as examples in a vocabulary learning app \cite{leong_putting_2024}, or the identity of a virtual instructor \cite{pataranutaporn_ai-generated_2021}, thereby increasing their motivation. Similarly, by (privately) applying augmented reality filters to one's audience, it is possible to reframe the situation and thereby reduce public speaking anxiety \cite{leong_picture_2023}. More provocative interventions can be imagined, such as gamifying cognitive effort, or leveraging the common feeling of being `unsettled' during decision making to provide timely provocations. Ethical implications emerge: for example, where does human agency fit in, and what is the responsibility of systems, particularly those that deliberately induce uncertainty, to provide emotional support to users? More fundamentally, what is the scope for GenAI-driven training, real-time support, or other approaches to augment `System 1' cognition, e.g., intuition \cite{boissin_bias_2021,chen_understanding_2023}, as an end in itself?

\subsection{Interacting and interfacing with AI} \label{subsec:interacting}
Behind the various approaches to augmenting cognition are considerations about how we interact and interface with AI.  Although turn-based conversational prompting has been a dominant interaction pattern for LLMs, other paradigms are emerging. Reviewing recent work, \ws{\citet{zindulka2025prompting}} argue that direct manipulation---prompting by manipulating relevant on-screen objects---can ease cognitive load associated with prompt formulation, more accurately express intent, and reduce the workflow disruptions that occur when navigating between content and the prompt interface. In turn, they argue that direct manipulation has the potential to free cognitive resources for higher-level thinking, and support the continuous expression of users' thought processes without being overwhelmed by AI output.

Other work has explored the dynamic between prompting and evaluating AI outputs. \ws{\citet{min_feedforward_2025}} propose `feedforward' information about the prospective AI output as a fundamental design component of GenAI systems. Alongside foreshadowing AI output, effective feedforward should enable users to efficiently disambiguate and reflect on prompts, and engage with feedforward content as a form of prompting. Through exploratory work, \citet{min_feedforward_2025} identify three initial dimensions for feedforward design: representation (e.g., topic lists), level of detail, and manipulability (e.g., selecting feedforward elements). These can be viewed as a form of `dynamic prompt middleware' \cite{drosos2025dynamic} for improving critical reflection during the prompting process.

Feedforward is one way to speed up the interactive loop between humans and AI. \ws{\citet{liu_interacting_2025}} approach this from a more radical angle, proposing a paradigm in which systems continuously generate internal and partly intrinsically-driven responses (`thoughts'), and selectively yet proactively communicate those responses to users throughout an interaction: full-duplex communication, in which proactive AI responses are gated by human heuristics for contributing to conversations. An instantiation of this paradigm shows promise in conversational engagement, and offers the possibility of a shared cognitive space in which both human and AI can build on fragments of ideas \cite{liu_proactive_2025}. 

Other interaction modalities are also being explored. For example, the recent rapid improvement in speech recognition and generation has led to increasing adoption of Voice User Interfaces \cite{seaborn_qualitative_2024}. Speech interaction may encourage more externalization of thoughts \cite{reicherts_its_2022,zavaleta_bernuy_does_2024}, whereas nonverbal aspects of speech like tone, loudness, and pauses can convey complementary information (e.g., emotional state \cite{seaborn_qualitative_2024}). What are other approaches to expressing human intent that have yet to be explored?         

How might different interaction approaches augment users' expression of intent, and might there be unexpected knock-on effects on cognition? For example, \citet{zindulka2025prompting} surmise that direct manipulation could also \textit{negatively} impact cognition. Given that writing \textit{is} thinking \cite{flower_cognitive_1981}, the replacement of text prompting with direct manipulation may remove opportunities for users to think through their intents and decompose their tasks (see also \cite{zindulka_content-driven_2025}). Moreover, the blending of user- and AI- generated content in the interface may encourage analogous blending in users' minds, including any potential biases of LLMs \cite{williams-ceci_biased_2025}. Likewise, how does the anthropomimetic slant of approaches like those in \citet{liu_interacting_2025} trade off in terms of improved conversational flow, users' mental models, over-reliance, emotional impact, and other dimensions? Future work should map the cognitive affordances of different modalities, including how they might express or transform knowledge and other internal states, and how such information might be represented back to users to augment cognition.

\subsection{Augmenting workflows} \label{subsec:augworkflows}

High-level intents rarely remain confined to discrete tasks. Rather, they ladder up to workflows (as per §\ref{subsec:workflows}). To truly augment cognition therefore involves thinking beyond the task to consider the broader workflows which encompass them \cite{palani_evolving_2024,tankelevitch_metacognitive_2024,vanukuru_designing_2025}. Workflows increasingly interleave manual and AI-assisted processes, human collaborators, and AI agents that coordinate to complete tasks. GenAI's flexibility and scale at handling context invite us to re-imagine workflows from first principles. 

One possibility is a move away from application-centric computing, where workflows are fragmented across siloed applications, to \textit{activity-centric computing} \cite{bardram_activity-centric_2019}, where application boundaries are dissolved in favor of coherently supporting users' broader workflow goals and activities. To this end, \ws{\citet{xia2025generative}} proposes a new AI-powered paradigm in which interfaces are dynamically generated to suit a user's task context, needs, and preferences (see also \cite{cao_generative_2025}). Their approach uses AI to infer the underlying task-driven models from users' prompts, representing the information entities, relationships, and data within information tasks. These task-driven models map to UI specifications, which are then used to dynamically generate interfaces suitable to users' tasks. Such approaches raise implications for cognition: for example, how can systems accurately determine users' higher- and lower- level goals across workflows, while minimizing the cognitive burden on users to correct inaccurate goal inferences and manually configure interfaces?    

Workflows extend not only across tasks, but also across people. The modern workplace is characterized by frequent collaboration across large networks. Yet, despite our inherently social nature, achieving efficient collaboration remains an ongoing challenge \cite{scott_mental_2024,yang_effects_2022}. How can AI support more effective communication and coordination of tasks? \ws{\citet{mukhopadhyay2025tailoring}} explore how AI could augment leadership behaviours for collaborative project development (see also \cite{mukhopadhyay_osint_2025}). Although their conversational AI system could support aspects of planning (e.g., by generating templates) or problem-solving (e.g., by providing feedback), they found that its impact “remained constrained to specific tasks rather than seamlessly integrating into the overall workflow”. Indeed, like most modern collaborative work, participants’ communication and workflows were distributed across interfaces which limited the AI system’s reach and contextual understanding \cite{yun_generative_2025}. Integrating AI support into realistic workflows is an open challenge, touching upon interaction design questions such as alternatives to the chat-based paradigm (§\ref{subsec:interacting}), and the role of proactive AI assistance \cite{liu_proactive_2025,deng_towards_2024,chen_are_2025, pu_assistance_2025, chen_need_2025, houtti_observe_2025, wang_social-rag_2025}. 

Not only does AI have potential to assist collaborative workflows, \ws{\citet{johnson2025promise}} propose the inverse: collaborative, multi-user interactions with AI can mitigate over-reliance observed with individual interactions. They posit that group interactions can encourage active engagement with AI assistance, for example, because the responsibility of critically evaluating AI contributions (and the impact of doing so) is distributed across members (e.g., see \cite{de_brito_duarte_amplifying_2025}). Groups' tendency for collaborative boundary regulation---the ``process of establishing limits and ground rules to maintain a healthy group environment''---can similarly regulate the influence of AI on group processes and outputs. Yet, AI may also plausibly exacerbate unproductive group dynamics like social loafing and group-think. Designing collaborative AI-assisted workflows to harness the benefits of both group work and AI assistance is an open challenge \cite{johnson_exploring_2025}. For example, which aspects of collaborative processes can be offloaded to AI and when? How can we detect and mitigate inappropriate reliance in group settings? How can we flexibly adapt AI's role in a group setting to suit different needs, and to shift between individual and group support? 

The rise of ``agentic'' AI---systems that attempt to execute workflows autonomously---raises further questions for collaborative work. If humans will increasingly hand over the production of code \cite{sarkar2025vibecoding, sarkar2023eupgenai}, writing \cite{dang_choice_2023}, and other tasks to autonomous agents, then how will human roles change? Will roles shift to management, including the definition of goals, outputs, guardrails, and verification procedures? How can people preserve and grow the expertise needed to assess the correctness of a workflow? As AI agents begin to work across people and teams, their management, including `context engineering' \cite{mei_survey_2025} and verification, will only increase in complexity. What will AI-augmented workflows look like in this context? How can AI support humans in the cognitive management of these complex workflows?

\section{Formative research, theory, measurement, and evaluation}
Understanding the impact of GenAI and designing systems that both protect and augment cognition will require formative research on GenAI users, workflows, and expertise, as well as a coherent approach and toolkit for running rigorous research studies, including theories, construct definitions and metrics \cite{wallach_evaluating_2024}, tasks, and protocols. Establishing a shared foundation and methodological approach will enable cross-study comparability, cumulative progress, and more intentional design of GenAI systems that support rather than undermine human thought.

\subsection{Formative research: Users, workflows, and expertise}\label{subsubsec:users}

We need a deeper understanding of distinct sets of GenAI tool users, their workflows, and how their expertise—in their domains, in GenAI, and in the management of the two (§\ref{subsubsec:expertise})—is instantiated within them. How are cognitive tasks and behaviors structured and sequenced? This is necessary to understand how GenAI changes or eliminates many of the micro-decisions people make in their workflows. To this end, \citet{nishal2025designing} explore the writing workflows of science journalists, whereas \citet{siu_augmenting_2025} describe the research survey practices of CS researchers, and the document review workflows of knowledge workers. Beyond work, \ws{\citet{kim_beyond_2025}} explore how `heavy' LLM users employ these tools for decision-making in their daily lives, defining such users as those who use LLMs for a substantial (and specifically defined) subset of concrete tasks that cover both work and personal domains, and spanning the range from purely rational to purely intuitive decision making, according to Cognitive Continuum Theory. Alongside the interview and survey methods deployed in these studies, \citet{schnizer2025user} propose that expertise can also be inferred via complementary approaches such as user interaction patterns, like prompting or eye-tracking. Without a clear understanding of who the ‘users’ are, any inferences and generalizations we make about the impact of AI will remain too vague or too muddled to be useful. Equally important, such understanding unlocks the ability to design systems tailored to users' expertise levels and needs \cite{schnizer2025user}. 

\subsection{Theories and constructs}\label{subsubsec:theory}
 What do we specifically mean by terms like “critical thinking”, “creativity”, “agency”, “learning”? Emerging research has drawn on a range of existing theories, models, and constructs from across disciplines: Bloom’s Taxonomy \cite{krathwohl_revision_2002}, Dewey’s theory of reflective thinking \cite{dewey_how_1910}, Schön’s model of reflection \cite{schon2017reflective}, Sawyer’s model of creativity \cite{sawyer_iterative_2021}, Hammond’s Cognitive Continuum Theory \cite{hammond2000human}, dual-process theories of cognition \cite{evans_dual-process_2013}, theories of metacognition \cite{tankelevitch_metacognitive_2024}, and pragmatism \cite{dewey_logic_1938}, among others. How far can these theories and associated constructs carry us today? Which theories and constructs are we missing? Do we need new or updated theories? For example, \ws{\citet{felten_beyond_2025}} argues that further theory development is needed to understand how human and AI biases can either amplify or diminish each other, i.e., moving away from separate treatment of these sources of bias and instead considering \textit{compound human-AI bias}. This interactionist perspective can help us develop strategies to mitigate the negative impacts of biases on human cognition.  
  
 \ws{\citet{slovak2025tools}} offers a concrete avenue for how theories can be used effectively. If GenAI systems are viewed as psychological interventions, then theories can inform \textit{theories-of-change} \cite{de_silva_theory_2014} for these interventions, which prescribe ``a particular set of experience trajectories that are expected to lead to the psychological effects for the participant(s)'' (see also \cite{slovak_hci_2024}). If theories and the resulting theories-of-change can prescribe (or describe) in detail what a user \textit{should} or \textit{would} experience when interacting with an AI system, then it becomes easier to translate these theories into design briefs that inform system design for cognitive augmentation, or, for existing systems, help uncover their positive or negative impact. 
  
\subsection{Measurement and evaluation}\label{subsubsec:eval}

Beyond defining constructs, how do we reliably measure them? Here, the disciplines with which HCI commonly intersects—cognitive, learning, information, and organizational sciences, and others—will become even more vital as sources of foundational research into cognition \cite{wallach_evaluating_2024,tankelevitch_metacognitive_2024}.

Yet, other disciplines may be insufficient on their own, as GenAI introduces new challenges and opportunities for evaluation. For example, human-AI interactions increasingly involve qualitative, open-ended tasks that may lack ground truth and often require iterative interactions, whereas much research in cognitive science (and human-AI interaction) has relied on simple, closed-ended tasks for experimental control. Therefore, there is a need for new methods to measure these interactions and task outcomes effectively. Prompting offers one such window into cognition, including users’ level of expertise, and their cognitive strategies, such as task decomposition \cite{kazemitabaar2024steering} and abstraction \cite{liu2023grounded,sarkar2022programmingai}, as well as their biases or gaps in understanding. These can be studied through user-centric approaches such as participatory prompting \cite{sarkar2023participatory, drosos2024bing}. They can also be approached in an automated manner, leveraging GenAI’s natural language capabilities. For example, \ws{\citet{holstein_consumption_2025}} present a promising novel framework and automated pipeline for analyzing prompting patterns along two dimensions: cognitive activity mode (exploration vs. exploitation) and cognitive engagement mode (constructive vs. detrimental).  

Another challenge is that many constructs central to this space, like critical thinking and agency, are inherently introspective and subjective. Collecting reliable self-report data on them requires research participants to have self-awareness, and even well-defined theoretical constructs may not align with people’s self-understanding. How do we measure something like “critical thinking” if people differ in their views of the term and of their own thinking? Despite the challenges, a turn to theory here can provide a starting point. For example, Dewey’s theory of reflective thought defines it as “definite units that are linked together so that there is a sustained movement to a common end” \cite{dewey_how_1933}. This quality of ‘traceability’ offers an intriguing avenue to characterising thinking with and without GenAI assistance. Similarly, Dervin’s sense-making theory \cite{naumer2008sense}, including the concept of cognitive gaps and bridges, offers another promising direction. These conceptualizations can inform the elicitation and analysis of people's varied experiences \cite{chen_understanding_2023}, which can now be conducted at scale using GenAI \cite{skeggs_micro-narratives_2025}. Ultimately, although the misalignment between theoretical definitions and subjective introspection may remain unresolved, it nevertheless deserves consideration in future work.

Finally, many potential changes in workflows and impacts on cognition may take time to emerge as users adopt tools and adapt their practices. How does prolonged use of GenAI affect users’ ability to reason independently, sustain attention, or engage in reflective thinking? Longitudinal studies are needed to track and understand potential changes.

\section{Conclusion}
GenAI poses a myriad of risks and opportunities for human cognition that present a large research and design space to address (and one that cannot be comprehensively addressed in one workshop). Building on the breadth of our workshop, we encourage future workshops and initiatives to take a focused approach, targeting specific areas to make concerted progress. We also encourage the HCI community to further expand its awareness of, and collaboration with, other communities and disciplines so as to share insights and accelerate progress. Finally, we emphasize that ongoing developments in task automation should not distract from the protection of core aspects of human cognition, and equally importantly, the innovation of approaches to augment and transform human cognition for the better.     

\begin{acks}
We thank all of those who submitted to and attended the workshop (and those that weren't able to make it in person) for their contributions to the lively discussions: Asif Ali, Ashton Anderson, Dinesh Ayyappan, Caroline Berger, Jessica Bo, Daniel Buschek, Yining "Rima" Cao, Chance Castañeda, Xiang Anthony Chen, Yan-Ying Chen, Man Lai Cheung, Parmit Chilana, Lydia Chilton, Kiroong Choe, Youjin Choi, Sadat Shams Chowdhury, Peter Dalsgaard, Nicholas Diakopoulos, Moritz Diener, Steven Drucker, Raymond Fok, Yue Fu, Avijit Ghosh, Frederic Gmeiner, Sven Goller, Zhitong Guan, Alexis Hiniker, Matthew Hong, Joshua Holstein, Kenneth Holstein, David Joyner, Harmanpreet Kaur, Malik Khadar, Anjali Khurana, Eunhye Kim, Markus Langer, Florian Lehmann, Joanne Leong, Wengxi Li, Can Liu, Xingyu Liu, Kurt Luther, Nikolas Martelaro, Sven Mayer, Bryan Min, Jessica Mindel, Vikram Mohanty, Anirban Mukhopadhyay, Muhammad Naeem, Syeda Masooma Naqvi, Sachita Nishal, Will Page, Kris Pilcher, Josh Pollock, James Prather, Chris Quintana, Rebecca Quintana, Brent Reeves, Leon Reicherts, Steven Rick, Arvind Satyanarayan, Kathrin Schnizer, Jannek Sekowski, Jinwook Seo, Anjali Singh, Alexa Siu, Petr Slovak, Hayden Stec, Philipp Spitzer, Sangho Suh, Karan Taneja, Esen K. Tütüncü, Jennifer Wortman Vaughan, Nick von Felten, Sitong Wang, Haijun Xia, Ryan Yen, Minju Yoo, JinYoung Yoo, Manqing Yu, Emma Zhuang, Tony Zhang, and Tim Zindulka. We also thank Sean Rintel, Richard Banks, Siân Lindley, Britta Burlin, Pratik Ghosh, Payod Panda, and others in the Tools for Thought team for their support on workshop development and execution. Finally, we thank the CHI Student Volunteers for their help during the day.
\end{acks}

\bibliographystyle{ACM-Reference-Format}
\bibliography{other_refs}

\appendix

\end{document}